\documentclass[a4paper,11pt]{article}
\pdfoutput=1 

\usepackage{jheppub} 
\usepackage{hyperref}
\usepackage[T1]{fontenc} 
\usepackage{bbm}
\usepackage{physics}
\usepackage[dvipsnames]{xcolor} 
\usepackage{wrapfig}
\usepackage{amsmath}
\DeclareMathOperator\arctanh{arctanh}
\usepackage{algorithm}
\usepackage[noend]{algpseudocode}
\usepackage{mathrsfs}
\usepackage{dsfont}
\usepackage{mathtools}
\usepackage{soul}
\usepackage{subcaption}
\usepackage[normalem]{ulem}

\newcommand{\footref}[1]{Footnote~\ref{#1}}

\title{\boldmath Holographic duality for Ising CFT with boundary}

\author[a]{Andreas Karch,}
\author[b]{Zhu-Xi Luo}
\author[a]{and Hao-Yu Sun}

\affiliation[a]{Department of Physics, University of Texas, Austin, TX 78712-1192, USA}
\affiliation[b]{Kavli Institute for Theoretical Physics, University of California, Santa Barbara, CA 93106-4030, USA}

\emailAdd{karcha@utexas.edu}
\emailAdd{zhuxi\_luo@kitp.ucsb.edu}
\emailAdd{hkdavidsun@utexas.edu}

\abstract{We extend the holographic duality between 3d pure gravity and the 2d Ising CFT proposed in Ref. \cite{castro2012gravity} to CFTs with boundaries. Besides the usual asymptotic boundary, the dual bulk spacetime now has a real cutoff, on which live branes with finite tension, giving Neumann boundary condition on the metric tensor. The strongly coupled bulk theory requires that we dress the well-known semiclassical AdS/BCFT answer with boundary gravitons, turning the partition function into the form of Virasoro characters. 
Using this duality, we relate the brane tensions to the modular $S$-matrix elements of the dual BCFT and derive the transformation between gravitational solutions with different brane tensions under modular $S$ action. 
}

\begin{document} 
\maketitle
\flushbottom

\section{Introduction}
\label{sec:intro}

One out of several problems in the study of pure AdS$_3$ gravity is the infinite Poincar\'e series arising from the summation over semiclassical saddle points from Maloney and Witten \cite{maloney2010quantum}, which requires extensive efforts to regularize. 
In the pioneering paper \cite{castro2012gravity}, the study was extended into the quantum regime where the Brown-Henneaux central charge \cite{Brown:1986nw} $c=3\ell_{AdS}/2G_N$ is of order one. The Virasoro minimal models are considered as the conformal field theories dual to  pure AdS$_3$ gravity.  Interestingly, the summation becomes finite at least for unitary cases. The main idea is as follows. From the foundational work by Brown and Henneaux \cite{Brown:1986nw} we know that there are two copies of the Virasoro algebra with central charge $c=3\ell_{AdS}/2G_N$ on the asymptotic AdS$_3$ boundary. This indicates that, under certain assumptions, the Virasoro characters of the corresponding CFT should be the result of a determination of the gravity partition function, which turns out to be given by the summation of modular images of a ``vacuum seed'', i.e., the vacuum conformal block of the dual CFT, over a finite-index mapping class group  representing the enhanced gauge symmetry at strong coupling. In particular, for $c=1/2$, one obtains a perfect match between the Ising CFT and the gravity partition functions, except for a mysterious proportionality constant, which persists in higher-genus in both handlebodies and nonhandlebodies \cite{Jian:2019ubz,nonhandle}. A similar match can be found for the tricritical Ising model, but is difficult to extend to  $c>7/10$.

One may wonder whether using the Brown-Henneaux central charge formula is still justified at small values of $c$, which will get both perturbative and nonperturbative corrections in $1/c$. For example, \cite{Cotler:2018zff} found that a one-loop correction to the Brown-Henneaux result shifts $c$ by $+13$. The exact matching of the partition functions found in \cite{castro2012gravity,Jian:2019ubz,nonhandle} seems to suggest that the duality using the naive central charge matching is indeed valid , i.e., all the corrections sum to zero.  Following previous literature, we assume the validity of the duality at $c=1/2$, which is supported by the calculations in this work.

One can wonder whether this equivalence of pure 3d gravity with $c=1/2$ and the Ising model extends to spacetimes with genuine boundaries. In order to support field theories with boundaries, the bulk theory has to be augmented with Randall-Sundrum (RS) branes \cite{Randall:1999vf}. It is known that RS branes with tensions can be used to smoothly truncate the bulk spacetime in order to yield the geometry dual to a CFT with a genuine boundary \cite{Karch:2000ct,Karch:2000gx,Azeyanagi:2007qj,Takayanagi:2011zk,Fujita:2011fp}. The brane introduces one more \textit{parameter} into the theory, the brane tension. More precisely, for every possible boundary condition, a different brane with different tension needs to be introduced. 
Assuming the duality holds between the Ising BCFT and the bulk gravitational theory with RS brane at $c=1/2$,  three different kinds of branes are allowed, corresponding to the three primary fields in the theory \cite{cardy1986effect}. The brane tensions can then be fixed by requiring that certain field theory quantities are correctly reproduced. They turn out to be related to the CFT boundary entropies first introduced in  \cite{affleck1991universal}.

In this paper, we propose an exact duality between the Ising model with boundaries and a pure AdS$_3$ bulk gravity with RS branes. We use the results for the CFT partition function on the disk, more precisely the boundary contribution to this partition function which is given by the boundary entropy, to fix the free parameter in the bulk: the brane tension. We then test the duality on cylinders, as tori truncated  by one or two RS branes. Here we use the leading-order results in the high and low temperatures to fix the asymptotic forms of the gravitational partition function. While we do not explicitly perform the path integral over the 2d boundary gravitons, the full partition function follows from the asymptotic form by Virasoro symmetry. As a byproduct of our analysis, we also obtain a prediction for the transformation properties of the gravitational partition function under modular $S$ transformations.

The organization of the rest of this paper is as follows. In section \ref{sec:review}, we review the works on establishing the duality between 3d pure Einstein gravity and 2d Ising CFT on torus \cite{castro2012gravity} and closed Riemann surfaces of arbitrary genus \cite{Jian:2019ubz}. We also summarize the semiclassical duality between AdS and boundary CFT (BCFT) and the partition function of BCFT on an annulus. In section \ref{sec:duality}, we compute the bulk partition function dual to the Ising CFT on the cylinder both in the high- and low- temperature limits and match with field theory expectations. This section will be self-contained with regard to conventions and notations, to facilitate experts who would like to skip the review section. In section \ref{sec:general} we show that much of the structure we discovered for the Ising model carries over to general minimal models. However, in some of these latter cases it is questionable whether there even exists a simple duality between the CFT and pure gravity. Our analysis simply shows that the general structure of the partition functions would allow an extension to BCFTs by inclusion of RS branes if ever a gravity dual for these cases would be found. By assuming the validity of the duality, we also make a prediction about how brane tensions would change under modular $S$ transformations.

\section{Brief review of relevant concepts}
\label{sec:review}

In this section, we provide a short review of the Ising/gravity duality on closed manifolds, a brief introduction to boundary conformal field theory (BCFT) as well as the AdS/BCFT duality in the semi-classical regime. 

\subsection{Review of the Ising/gravity duality}
In ref. \cite{castro2012gravity}, the exact duality between 2d Ising CFT and AdS$_3$ pure gravity on a hyperbolic solid torus with $c=3\ell_{AdS}/2G_N=1/2$ is proposed by matching their partition functions up to a proportionality constant. The quantum partition function in the bulk is obtained as a sum over semi-classical saddle points (which are related to each other via $SL(2,\mathbb{Z})$ modular transformations), dressed by fluctuations of dynamical 2d boundary gravitons. Since the theory is strongly coupled, there is no sense in which the latter sum can be done perturbatively. It is one of the key assumptions of the proposal of \cite{castro2012gravity} that this procedure indeed accounts for the full bulk partition function. 

The modular sum can be organized into a summation of the modular images of the gravitational partition function of the thermal AdS$_3$ saddle $Z_{vac}$ and the boundary graviton fluctuations around it. One new feature of the work of ref \cite{castro2012gravity} compared with the semi-classical case \cite{maloney2010quantum} is that the summation is no longer over the infinite coset space $\langle T\rangle\backslash SL(2,\mathbb{Z})$, where $\langle T\rangle$ is the translation subgroup of $SL(2,\mathbb{Z})$. $\langle T\rangle$ is generated by $T=\begin{pmatrix}1 & 1\\0&1\end{pmatrix}$, which shifts the modular parameter $\tau$ of the asymptotic boundary torus by one and preserves the thermal AdS$_3$ saddle in the semi-classical case. In the quantum regime, the summation is instead over $\Gamma_c\backslash SL(2,\mathbb{Z})$, where $\Gamma_c$, an finite-index subgroup of $SL(2,\mathbb{Z})$ \cite{Bantay}, is enlarged from $\langle T\rangle$ and leaves the quantum partition function $Z_{vac}$ invariant. The key player $Z_{vac}$ is also called the ``vacuum seed'', and is argued to be equal to $|\chi_{1,1}(\tau)|^2$, the modulus square of the Virasoro character of the identity primary of the Ising CFT.
The result of the summation yields the gravitational partition functions $Z_{\text{grav}}=8Z_{\text{Ising}}$. At this level, a similar duality can be proposed for the tricritical Ising model with $c=7/10$, where $Z_{\text{grav}}=48Z_{\text{tri-Ising}}$.

In ref. \cite{Jian:2019ubz}, the authors extended the duality to handlebodies.  It is shown using topological quantum field theory techniques that the gravitational partition function for $3\ell_{AdS}/2G_N=1/2$ is again equal to that of the Ising CFT up to an overall finite constant prefactor, which at genus two is $384$, namely $Z_{\text{grav}}=384Z_{\text{Ising}}$. Further extension to non-handlebodies is achieved in \cite{nonhandle}, where non-handlebodies are systematically constructed. In particular, for the twisted I-bundles  whose conformal boundary is of genus two, it is shown that $Z_{\text{TIB}}=96Z_{\text{Ising}}$.

\subsection{Review of BCFT}
$1+1$ dimensional CFTs on manifolds with boundaries were introduced by Cardy in \cite{cardy1986effect}. They can be defined with any number of boundary components. For one boundary component, the surface can be conformally mapped to the upper half-plane. 

More interesting is the case with two boundary components, e.g., a cylinder/annulus, where the conformally invariant boundary conditions $a$ and $b$ correspond to what we will call the Cardy states $|a\rangle$ and $|b\rangle$.  There is a one-to-one mapping between each of these allowed boundary conditions and each primary operator of the CFT. 

The cylinder is characterized by one dimensionless shape parameter $\delta$, the ratio of circumference of the circle over the length {$L$} of the cylinder. One can write down the corresponding partition function in either the closed (annulus) channel:
\begin{equation}
\label{eq:closed}
    Z_{ab}(\delta)=\langle a|e^{-\widehat{H}/\delta}|b\rangle,
\end{equation}
or in the open (cylinder) channel:
\begin{equation}
\label{eq:Zab_chi}
    Z_{ab}(\delta)=\Tr_{\mathcal{H}}e^{-\delta \widehat{H}_{ab}}=\Tr_{\mathcal{H}}q^{\widehat{L}_0-\pi c/24}=\sum_hn^{ab}_h\chi_h(q).
\end{equation}
The closed channel is interpreted as a theory on a circle of unit radius propagating over a finite Euclidean time $1/\delta$ between initial and final states characterized by $|a\rangle$ and $|b\rangle$. The open channel gives the finite temperature partition function of the theory on the unit interval with boundary conditions characterized by $a$ and $b$ and inverse temperature $\beta=\delta L$. In \eqref{eq:Zab_chi}, $\chi_h(q)$ is the Virasoro character of the \textit{irreducible} representation\footnote{As opposed to the generic Virasoro character $\chi_{c,h}(q)=\frac{q^{h+(1-c)/24}}{\eta(\tau)}$ of a possibly reducible representation, where Dedekind's eta function $\eta(q)$ is defined in \eqref{eq:dedekind}.} of highest weight $h$ (corresponds to the primary field of conformal dimension $h$) and $q=e^{-\pi\delta}$. The last equality is due to the decomposition of the Hilbert space into highest weight representations of the Virasoro algebra,
\begin{equation}
    \mathcal{H}=\bigoplus_{h,\bar{h}} n_{h,\bar{h}}\mathcal{V}_h\otimes \overline{\mathcal{V}}_{\bar{h}},
\label{eq:decompose}
\end{equation}
where $n_{h,\bar{h}}$ are integers. The rank-three tensor $n_h^{ab}$ in \eqref{eq:Zab_chi} also takes values in $\mathbb{Z}$, and their values will be derived below.

Since the presence of the boundary ties the holomorphic sector of the CFT to the anti-holomorphic sector only, the partition function is a linear, not bilinear, combination of Virasoro characters \cite{francesco2012conformal}. The dilatation operator is now $\widehat{L}_0$ instead of $\widehat{L}_0+\widehat{\overline{L}}_0$. To determine the boundary states, one imposes the constraint
\begin{equation}
\widehat{L}_n|B\rangle=\widehat{\overline{L}}_{-n}|B\rangle.
\end{equation}
Because one can decompose the Hilbert space into vector spaces associated with conformal primaries as in \eqref{eq:decompose}, we can define \textit{Ishibashi states} \cite{ishibashi1989boundary,onogi1989conformal}  as
\begin{equation}
    |h\rangle\rangle\equiv\sum^{\infty}_{N=0}\sum^{d_h(N)}_{j=1}|h,N;j\rangle\otimes\overline{|h,N;j\rangle},
\end{equation}
where $d_h(N)$ is the dimension of the subspace of $\mathcal{V}_h$ at level\footnote{Given $\ket{h}$ the highest-weight state such that $\widehat{L}_0\ket{h}=h\ket{h}$, $\widehat{L}_{-k_1}\dots \widehat{L}_{-k_n}\ket{h}$ ($0\leq k_1\leq\dots\leq k_n$) is an eigenstate of $\widehat{L}_0$ with eigenvalue $h+k_1+\dots+k_n$, and its \textit{level} is $N=$$\sum_{i=1}^n k_n$.} $N$, and $|h,N;j\rangle,1\leq j\leq d_h(N)$ is an orthonormal basis for $\mathcal{V}_h$.

The state $|a\rangle$ associated to boundary $a$ is a linear combination of Ishibashi states. By equating open-channel and closed-channel partition functions and using the modular $S$ transformation on Virasoro characters, one obtains the Cardy conditions on the allowed boundary states $|a\rangle$:
\begin{equation}
\label{eq:conditions}
    \begin{split}
        n_{ab}^h&=\sum_{h'}S^h_{h'}\langle a|h'\rangle\rangle\langle\langle h'|b\rangle,\\
        \langle a|h'\rangle\rangle\langle\langle h'|b\rangle&=\sum_hS^{h'}_hn^h_{ab}.
    \end{split}
\end{equation}
These highly constraining equations enjoy unique solutions, \textit{Cardy states}, for all diagonal Virasoro minimal models\footnote{For a non-diagonal Virasoro minimal model which can be written as diagonal one in an extended algebra, formulae are similar. For example in \cite{Cardy:1989ir}, 3-state Potts being diagonal in $\mathcal{W}_3$ algebra, enjoys a similar formula in terms of $\mathcal{W}$-Ishibashi states. The expression for general non-diagonal Virasoro minimal models is a nontrivial generalization though, see \cite{Behrend:1999bn}. For example, explicit Cardy states of the \textit{tricritical} 3-state Potts model $\mathcal{M}(7,6)$ are in \cite{Iino:2020ipa}.}:
\begin{equation}
    |h\rangle \equiv \sum_{h'}\frac{S^{h'}_{h}}{\sqrt{S_0^{h'}}}|h'\rangle\rangle.
 \label{eq:CardyState}
\end{equation}
For each primary $h$ in the CFT, there is a corresponding conformal boundary condition, i.e., allowed boundary state. Plugging \eqref{eq:CardyState} back into \eqref{eq:conditions}, one obtains
\begin{equation}
    n^h_{h'h''}=\sum_l\frac{S^h_lS^l_{h'}S^l_{h''}}{S^l_0},
\end{equation}
which is identical to the Verlinde formula \cite{Verlinde:1988sn} for the fusion coefficients $N^h_{h'h''}$, such that
\begin{equation}
\label{eq:fusion}
n^h_{h'h''}=N^h_{h'h''}.
\end{equation}
For a recent nice and compact review, see \cite{Cardy:2004hm} or chapter 11.3.2 of \cite{francesco2012conformal}.

Another important concept in BCFT is the \textit{boundary entropy}, or the logarithm of \textit{$g$-function}, first proposed by \cite{affleck1991universal}. In the thermodynamic limit where $\delta=\beta/L$ is very small, only ground states of $\widehat{H}$ contribute to \eqref{eq:closed}:
\begin{equation}
Z_{ab}\sim\langle a|0\rangle\langle0|b\rangle e^{\pi c/6\delta},
\end{equation}
and the thermodynamic entropy simplifies to
\begin{equation}
\label{universalform}
    S_{ab}\equiv \frac{\beta}{Z}\frac{\partial Z}{\partial\beta}-\ln Z=\frac{\pi c}{3\delta}+g_a+g_b,
\end{equation}
where $g_a\equiv \ln(\langle a|0\rangle)$ and $g_b\equiv \ln(\langle b|0\rangle)$ are called boundary entropies. 
As can be observed from this equation, the boundary entropies for different boundary components decouple. 
Similar to Zamolodchikov's $c$-function \cite{Zamolodchikov:1986gt} for the 2d bulk, boundary entropies are proved to be monotonic under RG flows \cite{Friedan:2003yc,Casini:2016fgb}. The boundary entropy is a property of a single boundary and can already be extracted from studying the theory on the disk.

\subsection{Review of AdS/BCFT}
One natural question is then: what is the possible gravitational theory dual to CFT living on bordered surfaces? According to the AdS/BCFT correspondence \cite{Karch:2000ct,Karch:2000gx,Azeyanagi:2007qj,Takayanagi:2011zk,Fujita:2011fp} the bulk dual to a BCFT is gravity with branes. In particular in \cite{Takayanagi:2011zk,Fujita:2011fp} Takayanagi and his collaborators calculated the semi-classical limit of gravitational saddle-point partition functions, and showed how to extract the boundary entropy of the dual BCFT, which we will review below.

The spacetime $N$ in AdS/BCFT has genuine boundareis $Q_a$ in addition to the usual asymptotic (conformal) boundary $M$. The dynamical degrees of freedom are Einstein gravity and branes that terminate the spacetime. The combined action is
\begin{equation}
\label{eq:action}
    I=I_G + I_Q=\frac{1}{16\pi G_N}\int_N\sqrt{-g}(R-2\Lambda)+\frac{1}{8\pi G_N}\, \sum_a {\int_{Q_a}}\sqrt{-h}(K-T_a),
\end{equation}
where $K=h^{\mu \nu}K_{\mu \nu}$ is the trace of the extrinsic curvature $K_{\mu \nu}=\nabla_{\mu} n_{\nu}$ for an outward unit normal vector $n$ to each boundary, and $R$ and $\Lambda<0$ are Ricci scalar and cosmological constant, respectively. We allowed for branes with different tensions $T_a$ corresponding to the different allowed boundary conditions in the CFT. Since these allowed boundary conditions correspond to the primary operators in the dual CFT, they are labeled by the same label $a$. In a given solution, not all allowed branes need to be present. In fact, since we will be only discussing disk and annulus, all our solution employ either one or two branes.

Away from the brane sources the equations of motion are just the vacuum Einstein's equation. The stress tensor of the brane $Q_a$ imposes
\begin{equation}
    K_{\mu \nu}-h_{\mu \nu}K=8\pi G_N T^{Q_a}_{\mu \nu},
\end{equation}
at the end of spacetime.
Here 
\begin{equation}
T^{Q_a \mu \nu}=\frac{2}{\sqrt{-h}}\frac{\delta I_{Q_a}}{\delta h_{\mu \nu}}.
\end{equation}

On an AdS$_d$ foliation of AdS$_{d+1}$
\begin{equation}
ds_{AdS_{d+1}}^2=d\rho^2+\cosh^2\frac{\rho}{\ell_{AdS}}ds^2_{AdS_d},
\end{equation}
where $-\infty<\rho<\infty$, the cutoff surface $Q_a$ is located at $\rho=\rho_*>0$, one finds that
\begin{equation}
\label{eq:universal}
    T_a=\frac{d-1}{d}K=\frac{d-1}{\ell_{AdS}}\tanh\frac{\rho^a_*}{\ell_{AdS}}
\end{equation}
which will be a universal formula. 

The simplest case of AdS/BCFT is the one where the brane $Q$ is topologically a disk and anchors on a circle on the asymptotic boundary $M$ in the upper half-space model. In fact, we can write down a bulk geometry of this type for every brane with tension $T_a$ in our theory, corresponding to the various disk partition functions with boundary conditions $a$ that one can evaluated in the dual BCFT. 
\cite{Takayanagi:2011zk} calculates the semi-classical partition function for the CFT on the disk.
From these semi-classical disk partition functions one can extract a relation  between the brane tensions and the corresponding boundary entropies:
\begin{equation}
\label{eq:bdryent}
g_a = \frac{\rho^a_*}{4 G} = \frac{c}{6}\arctanh{\frac{T_a \ell_{AdS}}{d-1}   } .
\end{equation}
The important point here is that in order to match the disk partition functions, the input parameters $T_a$ in our bulk action are completely determined. 

The next interesting topology is the annulus or cylinder. There are two geometries with branes that can realize a cylinder on the boundary.\footnote{In \cite{Takayanagi:2011zk} a third bulk solution, also with disconnected branes, was found that relies on standard Poincar\'e coordinates in AdS$_3$. In appendix \ref{disconnected} we show that this is, in fact, not a new configuration, but just the high temperature solution rewritten in different coordinates} From the bulk point of view, we want to study the system at finite temperature. There are two such solutions. Both are locally AdS$_3$ and their topology are solid tori. The first is thermal AdS$_3$ and is expected to give the dominant saddle at low temperatures. If we take the boundary cylinder to be realized by an interval in the non-compact $x$ direction times the circle parametrized by Euclidean time $\tau$ this metric reads 
\begin{equation}
\label{thermalads}
    ds^2=\frac{\ell_{AdS}^2}{z^2}\left(d\tau^2+\frac{dz^2}{h(z)}+h(z)
     dx^2 \right),
\end{equation}
where $h(z)=1-(z/z_0)^2$, {and $2\pi z_0$ is the spatial periodicity.}
In this geometry the cycle that is contractible in the bulk is the spatial $x$ direction. Correspondingly, the only way to have a consistent brane configuration in the bulk is to have a single brane smoothly connect the two boundaries of the interval, since the brane can not end in the bulk unless it is wrapping a vanishing cycle. As a consequence, this connected solution only exists when both boundaries are described by the same boundary condition, since one and the same brane connects both.

The explicit solution for this configuration has been found in \cite{Takayanagi:2011zk, Fujita:2011fp}, the exact form is not relevant here. The total action turns out to be
\begin{equation}
\label{eq:nobdry}
    I_E=-\frac{\pi c \delta}{24},
\end{equation}
Interestingly, the contribution from the boundary/brane $Q$ cancels and so the action for this solution is completely independent of the brane tension, or in other words it is completely independent on which boundary condition we chose, as long as it is the same on both boundaries of the cylinder. This feature is in agreement with the behavior of boundary entropy of a 2d CFT on a very long cylinder \cite{Affleck:1986bv}.

The second saddle is the BTZ black hole, dominant at high temperatures. The metric is given by
\begin{equation}
\label{btz}
    ds^2=\frac{\ell_{AdS}^2}{z^2}\left(f(z)d\tau^2+\frac{dz^2}{f(z)}+dx^2\right),
\end{equation}
where $f(z)=1-(z/z_H)^2$. This time the Euclidean time $\tau$ is compactified on a circle with period $2\pi z_H$ so the BCFT temperature is $\beta=2\pi z_H$. Note that $\tau$ cycle is contractible since $f(z)=0$ when $z=z_H$. So this time the branes wrapping the Euclidean time direction can smoothly end at $z=z_H$ and we can have a disconnected configuration with different branes ending on the two ends of the interval. Correspondingly, this configuration is allowed even when we impose different boundary conditions on the two ends of the interval. The branes $Q_1$ with tension $T_a$ and $Q_2$ with tension $T_b$ are perpendicular to the spatial slice and separated by $\Delta x=\pi z_0$. In this case the total action turns out to be
\begin{equation}
\label{eq:BTZaction}
    I^{ab}_E=I^a_{\text{bdry}} + I^b_{\text{bdry}}+I_{\text{bulk}}=-\frac{c}{6}\left ( \arctanh(\ell_{AdS}T_a ) + \arctanh(\ell_{AdS} T_b) \right ) -\frac{\pi c}{6\delta}.
\end{equation}
Note that this expression agrees with the universal form \eqref{universalform}.

In the holographic setting of \cite{Takayanagi:2011zk} there is a sharp first-order phase transition between the two saddles. This is an artefact of the large central charge limit. We already saw in the previous section that the full cylinder partition function is a smooth function of $\delta$. So the two saddles will always both contribute at finite central charge.\footnote{Strictly speaking, modular invariance always forces us to sum over all saddles. However, in the large central charge $c$ limit, each saddle contributes as $e^{- c S_s} $, where $S_s$ encodes the action of the saddle. Consequently, what appears as a smooth function at finite $c$ from the sum over saddles, at large $c$ degenerates into a non-differentiable function that is dominated by a single saddle for each set of parameters.}
However, as we will see in detail, the leading contribution at high and low temperatures respectively will come from the corresponding saddles. Furthermore, working at $c=1/2$ means that we will have to move beyond the simple semi-classic answer of \cite{Takayanagi:2011zk} and account for quantum fluctuations. We will argue that the form of these fluctuations is completely fixed by the conformal symmetry of the problem.

\section{Duality for the Ising BCFT}
\label{sec:duality}
In this section, we will match the semi-classical gravity partition functions in \cite{Takayanagi:2011zk} with the leading terms of the BCFT results at different limits. We will focus on the case where the Brown-Henneaux central charge is $1/2$, such that the corresponding BCFT is the Ising theory. The following notations will be used: $1,\psi,\sigma$ label the three primary fields in the 2d Ising CFT with conformal dimensions $0,1/2, 1/16$, respectively. Since there is a bijection between the Cardy states and the bulk primary fields \cite{Cardy:2004hm}, we will denote the boundary states using $1,\psi,\sigma$ as well. 

\subsection{Disk Partition Function}
\label{sec:disk}

As reviewed in section \ref{sec:review}, in the semi-classical regime, the AdS/BCFT duality suggests a relationship between the brane tension and boundary entropy \eqref{eq:bdryent}. This relation can already be extracted from studying the theory on the disk. If our conjectured duality is true in the quantum regime of $c=3\ell_{AdS}/2G_N<1$, then a similar relation should hold. In particular, for the half-space model the with a single brane, we expect the boundary entropy to be
\begin{equation}
g_a=\frac{c}{6}\arctanh(\ell_{AdS} T_a).
\label{eq:main}
\end{equation}
Here $\ell_{AdS}$ is the AdS radius which we fix to be $1$ from now on, $T_a$ is the RS brane tension for boundary condition $a$, and $\log g_a$ is the boundary entropy in the BCFT with boundary condition $a$. In particular for $c=1/2$, we have
\begin{equation}
    T_1=T_\psi=-\tanh (6\log 2),\quad T_\sigma=0.
\label{eq:tensions}
\end{equation}
The boundary state associate with $\sigma$ corresponds brane to a tensionless brane. The other two tensions are degenerate and negative. 

Negative tension branes are problematic if we think of them as fluctuating objects. The worldvolume scalar (or radion) representing transverse motion of the brane would have a negative kinetic term; the energy of the brane would be unbounded from below as the surface of the brane becomes arbitrarily rough. The way to avoid this is to declare the brane not to be an fluctuating object, but merely is a fixed defect that carries energy density which can deform the spacetime around it. Examples of such defects are orbifolds in string theory, which are fixed planes of a symmetry projection that removes states from the spectrum odd under the symmetry. It is quite common to require RS branes to be non-fluctuating orbifolds (as in the original RS1 model \cite{Randall:1999ee}), so this is not an unreasonable solution. While the zero tension brane (and the positive tension branes we will find in the tricritical Ising generalization) could in principle be fluctuating objects, self-consistency of the constructions seems to require to treat all branes on an equal footing and so we will treat all branes as not fluctuating.

\subsection{Cylinder in the low-temperature limit}
In the low-temperature limit, the dominant solution to the Einstein equation is that of the thermal AdS$_3$ (shorthand notation TAdS). Since we defined $\delta$ as the ratio between the circumference and the height of the cylinder\footnote{Our $\delta$ here is same as $1/\left(\Delta x\cdot T_{BCFT}\right)$ in \cite{Fujita:2011fp,Takayanagi:2011zk}. \label{foot:delta}}, the low-temperature case corresponds to large $\delta$. In  \cite{Takayanagi:2011zk}, the leading contribution to the TAdS partition function is found in the semi-classical regime to be
\begin{equation}
Z^{\text{TAdS}}=e^{\pi\delta/48}+{\dots}=q^{-1/48}+{\dots},
\label{eq:Ising_TAdS}
\end{equation}
where in the second equality we have parametrized $q=e^{-\pi\delta}\in\mathbb{R}.$

Now we turn to the BCFT perspective. For convenience, let us reproduce \eqref{eq:Zab_chi}: 
\begin{equation}
Z_{ab}(q)=\sum_h n_{ab}^h\chi_h(q),\nonumber
\end{equation}
namely, all $Z_{ab}$'s are linear combinations of different characters. 
In the low-temperature limit, $q$ is small and the leading term in the series expansion for $\chi_h(q)$ is $q^{h-1/48}$, where $h$ is the conformal weight of the primary field $h$. Since $h$ is non-negative, the contribution of $\chi_1(q)$ is always dominant in the summation above whenever it is present. Consequently, we will be interested in the $Z_{ab}$'s such that $n_{ab}^0$ is nonzero. This only happens when $a=b$:
\begin{equation}
Z_{ab}^{\text{low}}=\delta_{ab}~\chi_1(q)+\mathcal{O}(q^{23/48}) =
\delta_{ab}~q^{-1/48}+\mathcal{O}(q^{23/48}).
\label{eq:Ising_low}
\end{equation}

Equations \eqref{eq:Ising_TAdS} and \eqref{eq:Ising_low} exhibit a match. It is thus attempting to conjecture that the match extends beyond the low-temperature limit: at Brown-Henneaux central charge $c=3\ell_{AdS}/2G_N=1/2$, the gravitational partition function for a cylinder with two branes is equal to the BCFT partition function on the cylinder at the same central charge. In other words, there is AdS/BCFT duality in the quantum regime of $c=1/2$. The appearance of $\delta_{ab}$ in \eqref{eq:Ising_low} also allows for a gravitational interpretation: For the TAdS saddle on a solid torus, Euclidean time is along the longitude direction, i.e., the constant time slices are disks. Since the RS brane cuts through a constant time slice, we are led to a bagel-like cut of the torus into two annuli, and the boundary conditions for corresponding boundaries of both annuli must be the same. We will revisit this claim in section \ref{sec:3.4} after the high-temperature discussions.

We note that the arguments in this subsection can be straightforwardly generalized to all diagonal Virasoro minimal models, which will be discussed in section \ref{sec:general}.

\subsection{Cylinder in high-temperature limit}

To completely pin down the partition functions for different boundary conditions, we move on to the case of a cylinder with general (meaning not necessarily equal) boundary conditions, and study the high-temperature limit where the BTZ black hole solutions are favored. Ref. \cite{Takayanagi:2011zk,Fujita:2011fp} computed the case where the two branes on the cylinder have the same tension. The partition function separates into three decoupled contributions from the bulk and two boundaries. Slight generalization of their results leads to 
\begin{equation}
Z^\text{{BTZ}}_{ab}=e^{g_{ab}}~e^{\pi/12\delta}+\dots\equiv e^{g_{ab}}\tilde{q}^{-1/48}+\dots,
\label{eq:BTZ}
\end{equation}
where $\tilde{q}=e^{-4\pi/\delta}$, and it is small when $\delta$ is small (same as in \footref{foot:delta}). The constants $g_{ab}$ satisfy
\begin{equation}
g_{ab}=g_a + g_b,
\end{equation}
and $g_a$, $g_b$ are as in equation \eqref{eq:main}. 

On the BCFT side, since $q$ is close to $1$ from below, it is no longer justified to take the leading term in the $q$-expansion for the characters. Instead, we perform the modular $S$ transformation and expand in terms of $\tilde{q}$:
\begin{equation}
\chi_h(q)=\sum_{h'}S_{hh'}\chi_{h'}(\tilde{q})=  S_{h1} \tilde{q}^{-1/48} +\mathcal{O}(\tilde{q}^{h-1/48}).
\end{equation}
Recall that in the order of basis $\{1,\psi,\sigma\}$, the $S$-matrix of the Ising theory is
\begin{equation}
    S=\frac{1}{2}\left(\begin{matrix} 1 & 1 & \sqrt{2}\\ 1 & 1 & -\sqrt{2} \\ \sqrt{2} & -\sqrt{2} & 0 \end{matrix}\right).
\end{equation}
This leads to, at leading order, 
\begin{equation}
\chi_1\sim \frac{1}{2} \tilde{q}^{-1/48},\quad
\chi_\psi\sim \frac{1}{2} \tilde{q}^{-1/48},\quad 
\chi_\sigma\sim \frac{1}{\sqrt{2}}\tilde{q}^{-1/48}.\\
\end{equation}
The corrections are of order $\mathcal{O}(\tilde{q}^{-23/48})$.
Using \eqref{eq:Zab_chi}, the partition functions are then of the form
\begin{equation}
Z_{ab}^{\text{high}}=e^{g_{ab}} \tilde{q}^{-1/48}+\mathcal{O}(\tilde{q}^{-23/48}),
\end{equation}
where again $g_{ab}=g_a+g_b$ and
\begin{equation}
{g}_1={g}_\psi=-\log\sqrt{2},\quad {g}_\sigma=0. 
\label{eq:g1ss}
\end{equation}
Comparing with the gravity calculations, we observe a perfect match if the constants $g_{ab}$ in equation \eqref{eq:BTZ}, factorized as $g_{ab}=g_a + g_b$, are identified with the boundary entropies of the BCFT \eqref{eq:g1ss}. This serves as a nontrivial check of our starting point \eqref{eq:main} in the disk case.

\subsection{The full quantum partition function}
\label{sec:3.4}

So far we have argued that the AdS/BCFT calculations reproduce correctly the high and low temperature limits of the BCFT partition function $Z_{ab}(\delta)$. In order to reproduce the full partition function, we have to sum over fluctuations around these semiclassical saddles arising from boundary gravitons.
That is, the full quantum mechanical partition function is a double sum over all classical saddles, and for each saddle the weight is obtained not just from the action of the classical saddle itself, but also from summing up the contribution from all boundary graviton fluctuations around it. In the limit of a large central charge, the action of the saddles itself is of order $c$, so the subdominant saddle is exponentially (in $c$) suppressed. At any given set of parameters, only the contributions from high and low temperatures matter. 
As we mentioned above, this gives rise to the sharp phase transitions familiar from holographic studies \cite{Fujita:2011fp}.
Furthermore, the large central charge suppresses fluctuations including those from boundary gravitons. In our case, since $c=1/2$, we never get such a sharp transition. We should always sum over all saddles and include all fluctuations.
While it would be interesting to explicitly work out the contribution of these fluctuations, we note that the general form of the partition function is fixed by the Virasoro symmetries of the problem -- \eqref{eq:Zab_chi} is fixed by the underlying conformal invariance. In particular, the full $a$ dependence has to appear via characters $\chi_h(q)$. The only dynamical information are the coefficients $n^h_{ab}$. We can fix these coefficients from our results obtained in the extremely high and low temperature limits, where the semiclassical analysis in terms of high- and low-temperature saddles applies. Our low and high temperature calculations show that we reproduce the right structure and fix these coefficients (to be $0$ or $1$) from the saddle point analysis alone. The sum over saddles together with the boundary gravitons then, by symmetry, will have to reproduce the full Ising partition function. We here emphasize that we always assume RS branes are free of 2d boundary gravitons (small diffeomorphisms), consistent with the non-fluctuating nature of branes as stated near the end of section \ref{sec:disk}.

\section{General Virasoro minimal models}
\label{sec:general}

In this section, we turn to the general correspondence between diagonal minimal model BCFTs and pure gravity in AdS$_3$. A duality was only conjectured to be true in \cite{castro2012gravity} for the Ising and the tricritical Ising model, with the latter running into difficulties at higher genus \cite{Jian:2019ubz}. Here we will show that \textit{if} a theory of gravity can be found to reproduce the partition function of the Virasoro minimal model on any closed surface, the extension of the duality between a gravitational bulk with RS branes and BCFT on the asymptotic boundary is straightforward. 
In particular, a general formula relating the brane tension with the modular $S$-matrix in the corresponding BCFT will be presented and the case of tricritical Ising will be studied in detail. 

\subsection{Low-temperature limit}

For a general Brown-Henneaux central charge $c<1$, the leading low-temperature contribution to the thermal AdS$_3$ partition function is now \cite{Takayanagi:2011zk}, 
\begin{equation}
Z^{\text{TAdS}}=q^{-c/24}+{\dots},
\label{eq:TAdS}
\end{equation}
where again $q=e^{-\pi\delta}.$ The leading term in the small-$q$ series expansion for $\chi_h(q)$ is $q^{h-c/24}$. Since $h$ is non-negative, we still find that the contribution of the vacuum block $\chi_1(q)$ with $h_1=0$ is always dominant whenever it is present in the summation. Consequently, we will be interested in the $Z_{ab}$'s such that $n_{ab}^0$ is nonzero. This only happens when $a=b$:
\begin{equation}
Z_{ab}^{\text{low}}=\delta_{ab}~\chi_1(q)+\mathcal{O}(q^{h_*-c/24}) =
\delta_{ab}~q^{-c/24}+\mathcal{O}(q^{h_*-c/24}).
\label{eq:BCFT_low}
\end{equation}
Here $h_*$ is the smallest positive conformal weight in the CFT.

The match between equations \eqref{eq:TAdS} and \eqref{eq:BCFT_low} suggests that if a duality between a theory of gravity and a diagonal minimal model\footnote{The assumption of being diagonal under the \textit{Virasaro} algebra is important to obtain the form \eqref{eq:Zab_chi}.} with the same central charge can be found, the bulk can be augmented to be a BCFT by the inclusion of the appropriate RS branes.

\subsection{High-temperature limit}

In the high-temperature limit, in a BCFT with central charge $c$, the series expansion of $\tilde{q}$ gives:
\begin{equation}
\chi_h(q)=S_{h1} \tilde{q}^{-c/24} +\mathcal{O}(\tilde{q}^{h_*-c/24}).
\end{equation}
This leads to the following partition functions 
\begin{equation}
\label{eq:general}
Z_{ab}^{\text{high}}=e^{g_{ab}} \tilde{q}^{-c/24}+\mathcal{O}(\tilde{q}^{h_*-c/24}),
\end{equation}
where again ${g}_{ab}={g}_a+{g}_b$ and \cite{affleck1991universal,Affleck:1986bv}
\begin{equation}
{g}_a = {\log~} (S_{0a}/\sqrt{S_{00}})= {\log~} (d_a/\sqrt{D}).
\end{equation}
Here $d_a=S_{0a}/S_{00}$ is the quantum dimension for $a$ and $D=\sqrt{\sum_a d_a^2}$ is the total quantum dimension. Compared with the gravitational computation in \cite{Takayanagi:2011zk}, we identify
\begin{equation}
T_a=\tanh (\frac{6}{c}\log \frac{d_a}{\sqrt{D}} ).
\end{equation}
This relation can also be derived once again from the disk. 

\subsection{Modular \texorpdfstring{$S$}{} transformation}
In this part, we will predict the transformation of the gravitational partition function under the modular $S$-action based on the conjectured duality with BCFT.

Under the $SL(2,\mathbb{Z})$ transformation $S$, the BCFT partition functions change as
\begin{equation}
Z_{ab}(S\cdot q)=\sum_{h} n^{h}_{ab}~\chi_{h}(S \cdot q)=\sum_{h,h'} n^{h}_{ab}~ S^{h'}_h\chi_{h'}(q).
\label{eq:Z_S}
\end{equation}
In the first two expressions, the $S$'s are in the two-dimensional fundamental representation, while in the last expression, $S$ is in the conformal block basis. Alternatively, the transformed partition function can be written as a linear combination of various $Z_{cd}$'s with different boundary conditions:
\begin{equation}
Z_{ab}(S\cdot q)=\sum_{c,d}f^{cd}_{ab}~Z_{cd}(q).
\label{eq:f_S}
\end{equation}
These coefficients $f$'s have the following form
\begin{equation}
f^{cd}_{ab}=\sum_{h,h'}n^h_{ab}(S_{0h})^2S^{h'}_h n^{h'}_{cd}.
\label{eq:f}
\end{equation}
One can show that \eqref{eq:f}, when plugged into \eqref{eq:f_S}, reduces to \eqref{eq:Z_S}. The details are presented in appendix \ref{app:S}.

Equation \eqref{eq:f} describe the transformation properties of the gravitational partition function under modular $S$ action, which is a new result of the duality. The other generator of $T$ of $SL(2,\mathbb{Z})$, however, is not well-defined in the case with branes, as there is only one real parameter $q$ in the partition function, instead of two in the solid torus case. Hence, unlike in in \cite{Dijkgraaf:2000fq, maloney2010quantum}, there is no ``summation over geometry'' over $\mathbb{Z}$, the mapping class group of an annulus/cylinder.

The relation \eqref{eq:f_S} is interesting since it states that upon performing an $S$ transformation in the bulk of a gravitational solution with a given set of branes, the result arises as a non-trivial sum over several configurations employing different branes.

\subsection{An example: the Tricritical Ising model}
Now let us look at a specific example other than Ising CFT. The next simplest unitary minimal model is tricritical Ising. Its six primary fields are
\begin{equation}
\label{eq:tribasis}
    \chi_{1},\quad\chi_{\epsilon},\quad\chi_{\epsilon'},\quad\chi_{\epsilon''},\quad\chi_{\sigma},\quad\chi_{\sigma'},
\end{equation}
with conformal dimensions:
\begin{equation}
    0,\quad\frac{1}{10},\quad \frac{3}{5},\quad\frac{3}{2},\quad\frac{3}{80},\quad\frac{7}{16},
\end{equation}
respectively. The complete set of fusions rules is listed in \cite{francesco2012conformal} and all the fusion coefficients are either $0$ or $1$, which can be seen from the Verlinde formula and the fact \eqref{eq:fusion}. Explicit form of characters written in terms of generalized theta functions are written down in appendix \ref{app:tri}. 

In ref. \cite{castro2012gravity}, the authors showed that Ising and tricritical Ising CFTs are possible dual to the bulk with corresponding AdS radii $\ell_{AdS}$, as these are the only two theories with unique modular invariants. Further in \cite{Jian:2019ubz}, for an arbitrary higher-genus asymptotic boundary, the partition function of the bulk which is presumably dual to the tricritical Ising theory results in an infinite number of summands in the modular sum. The appropriate regularization over the mapping class group of $\Sigma_{g,0}$ with $g>1$ is unknown \cite{Jian:2019ubz,Sun:2020mee}. 
However, here we only focus on the case where the parent spacetime is a solid torus.

\subsubsection{The low-temperature limit}
At low temperature where the parent spacetime is the thermal AdS$_3$, again we have same boundary conditions on both ``sides'' of a single brane performing the ``bagel cut''. According to \eqref{eq:tricritical}, in this limit all possible annulus partition functions with same boundary conditions are the same to the leading order:
\begin{equation}
    \begin{split}
        Z_{\epsilon\epsilon}=Z_{\epsilon'\epsilon'}\sim\chi_1&\sim q^{-7/240}=e^{7\pi\delta/240},\\
        Z_{\epsilon''\epsilon''}=Z_{11}=\chi_1&\sim q^{-7/240},\\
        Z_{\sigma\sigma}\sim\chi_1&\sim q^{-7/240},\\
        Z_{\sigma'\sigma'}\sim\chi_1&\sim q^{-7/240}.\\
    \end{split}
\end{equation}
They agrees with the semi-classical result of $I_{\text{bulk}}$ as in \eqref{eq:nobdry} of \cite{Fujita:2011fp}, with Brown-Henneaux central charge $c=7/10$. Again, this suggests a potential duality between AdS$_3$ and tricritical Ising CFT. We now further investigate this possibility.

\subsubsection{The high-temperature limit}
In this the high-temperature case, the parent bulk is a BTZ black hole. Because now $q$ is large, in order to expand the proposed partition functions in terms of $q$, we perform a modular $S$ transformation again. Now $\tilde{q}=e^{-4\pi/\delta}$ is small when $\delta$ is small, and to the leading order we have
\begin{equation}
\chi_h(q)=\sum_{h'}S_{hh'}\chi_{h'}(\tilde{q})\sim \tilde{q}^{-c/24} (S_{h0}+S_{h\epsilon}\tilde{q}^{1/10}+S_{h\epsilon'}\tilde{q}^{3/5}+S_{h\epsilon''}\tilde{q}^{3/2}+S_{h\sigma}\tilde{q}^{3/80}+S_{h\sigma'}\tilde{q}^{7/16}).
\end{equation}
So at leading order:
\begin{equation}
    \begin{split}
        \chi_1\sim\chi_{\epsilon''}&\sim\frac{s_2}{\sqrt{5}}\tilde{q}^{-c/24}=\frac{s_2}{\sqrt{5}}e^{7\pi/30\delta},\\
        \chi_{\epsilon}\sim\chi_{\epsilon'}&\sim\frac{s_1}{\sqrt{5}}\tilde{q}^{-c/24}=\frac{s_1}{\sqrt{5}}e^{7\pi/30\delta},\\
        \chi_{\sigma}&\sim\frac{\sqrt{2}s_1}{\sqrt{5}}\tilde{q}^{-c/24}=\frac{\sqrt{2}s_1}{\sqrt{5}}e^{7\pi/30\delta},\\
        \chi_{\sigma'}&\sim\frac{\sqrt{2}s_2}{\sqrt{5}}\tilde{q}^{-c/24}=\frac{\sqrt{2}s_2}{\sqrt{5}}e^{7\pi/30\delta}.
    \end{split}
\end{equation}
All the possible partition functions in \eqref{eq:tricritical} are now
\begin{equation}
    \begin{split}
        Z_{\epsilon\epsilon}=Z_{\epsilon'\epsilon'}=Z_{\epsilon\epsilon'}\sim&\frac{s_1+s_2}{\sqrt{5}}e^{7\pi/30\delta},\\
        Z_{\epsilon\epsilon''}=Z_{1\epsilon'}=Z_{\epsilon'\epsilon''}=Z_{1\epsilon}\sim&\frac{s_1}{\sqrt{5}}e^{7\pi/30\delta},\\
        Z_{\epsilon''\epsilon''}=Z_{11}\sim Z_{1\epsilon''}\sim&\frac{s_2}{\sqrt{5}}e^{7\pi/30\delta},\\
        Z_{\epsilon\sigma}=Z_{\epsilon'\sigma}\sim&\sqrt{\frac{2}{5}}(s_1+s_2)e^{7\pi/30\delta},\\
        Z_{\epsilon\sigma'}=Z_{\epsilon'\sigma'}=Z_{\epsilon''\sigma}=Z_{\epsilon''\sigma'}=Z_{0\sigma}\sim&\frac{\sqrt{2}s_1}{\sqrt{5}}e^{7\pi/30\delta},\\
        Z_{\sigma\sigma}\sim&\frac{2s_1+2s_2}{\sqrt{5}}e^{7\pi/30\delta},\\
        Z_{\sigma\sigma'}\sim&\frac{2s_1}{\sqrt{5}}e^{7\pi/30\delta},\\
        Z_{\sigma'\sigma'}\sim&\frac{2s_2}{\sqrt{5}}e^{7\pi/30\delta},\\
        Z_{1\sigma'}\sim&\frac{\sqrt{2}s_2}{\sqrt{5}}e^{7\pi/30\delta},
    \end{split}
\end{equation}
where $s_1$ and $s_2$ are entries of the modular $S$-matrix reviewed in appendix \ref{app:tri}.
According to our proposition \eqref{eq:general} on the form of leading term in partition functions, gravitational $g_a$'s can be solved by boundary entropies of BCFT:
\begin{equation}
\begin{split}
    g_{{\epsilon}}=g_{\epsilon'}={\frac{1}{4}\log}\left(\frac{1}{4}+\frac{1}{2\sqrt{5}}\right),\\
    g_1=g_{\epsilon''}={\frac{1}{4}\log}\left(\frac{1}{8}-\frac{1}{8\sqrt{5}}\right),\\
    g_{\sigma'}={\frac{1}{4}\log}\left(\frac{1}{2}-\frac{1}{2\sqrt{5}}\right),\\
    g_{\sigma}={\log(2\sqrt{2}\cos\left(\frac{\pi}{5}\right))+g_1},
    \end{split}
\end{equation}
which exactly agree with the already known values of boundary entropy for the purely 2d tricritical Ising CFT \cite{Dorey:2009vg}. The four different RS brane tensions can be easily computed using \eqref{eq:main}.

\section{Discussion}
In this work, we demonstrated that the duality between pure gravity in AdS$_3$ and the Ising CFT can be extended to conformal field theories with boundaries. There are several interesting questions that should be addressed in the future.

For one, one could wonder whether an extension to higher central charges can be found. Since in AdS/BCFT, there is no Poincar\'e series, but only two terms, upon summing over geometries, we do not encounter the difficulty of regularizing an infinite sum. Hence, in principle, any rational CFT with $c>1$ (even irrational CFTs, as long as the modular $S$-matrix is well-defined, such as Liouville field theory \cite{McGough:2013gka} and logarithmic CFT \cite{Ridout:2014yfa}) would allow branes with certain finite tensions extending into the bulk. Of course this is a moot point unless a dual for the CFTs on closed manifolds can be constructed, but it is encouraging to note that no new constraints arise from the case with boundary. The last fact is expected, because in BCFT, only a $\mathbb{Z}_2$ subgroup (generated by $S$) of the original $SL(2,\mathbb{Z})$ modular symmetry is preserved, so we get less restrictive constraints. 

We only studied the case of the CFT partition function on a disk and a cylinder. A general orientable 2d manifold $\Sigma_{g,b}$ with boundaries is characterized by two nonnegative integers, the genus $g$ and the number of boundaries $b$. Our two cases correspond to $(g,b)=(0,1)$ and $(0,2)$. Extensions to higher $g$ and $b$ would be interesting. These more complicated geometries might allow us to connect to recent work on the gravitational determination of CFT correlation functions, which have been studied in e.g., \cite{Maloney:2016kee}.

It would be also interesting to see if our analysis can be generalized to interfaces connecting several CFTs rather than just boundaries. Such interface CFTs have a much richer structure of allowed boundary conditions and it would be illuminating to see if they can be reproduced from a putative holographic dual.

Last but not least, it would be interesting to explicitly perform the path integral over boundary gravitons, possibly by using a modified version of the heat kernel in \cite{Giombi:2008vd} on 3-manifolds with genuine boundary components set by RS branes. We deduced the result based on the structure of the partition function that is forced upon us by the conformal symmetry of the problem, but we would presumably learn something about quantum gravity in the presence of RS branes if we were to be able to redo this calculation explicitly in the gravity theory.

\acknowledgments
We thank Tadashi Takayanagi for helpful discussions, and Kristan Jensen and Alexander Maloney for providing comments on the draft. H.-Y. S. thanks Stephen Ebert, Dongsheng Ge, Hao Geng for useful discussions. All three authors are supported by the Simons Collaborations on Ultra-Quantum Matter, grant 651440 (AK and LB) from the Simons Foundation.

\appendix

\section{Proof of equation \texorpdfstring{\eqref{eq:f}}{}}
\label{app:S}

In this appendix, we show that \eqref{eq:f}, combined with \eqref{eq:f_S}, gives \eqref{eq:Z_S}.
\begin{equation}
\begin{split}
\sum_{c,d}f^{cd}_{ab}~Z_{cd} & = \sum_{c,d} \sum_{h,h'}n^h_{ab}(S_{0h})^2S^{h'}_h n^{h'}_{cd}~Z_{cd} \\
& = \sum_{h,h'}n^h_{ab}(S_{0h})^2S^{h'}_h\sum_{h''}\chi_{h''} \sum_{c,d} (n^{h'}_{cd} n^{h''}_{cd})\\
& = \sum_{h,h'}n^h_{ab}(S_{0h})^2S^{h'}_h\sum_{h''}\chi_{h''} \sum_{c,d}\sum_{x,y}S_{xc}S_{xd}S_{xh'}S_{yc}S_{yd}S_{yh''}/S_{x0}S_{y0}\\& = \sum_{h,h'}n^h_{ab}(S_{0h})^2S^{h'}_h\sum_{h''}\chi_{h''} \sum_{x,y} S_{xh'}S_{yh''} (\sum_{c}S_{xc}S_{yc})(\sum_{d} S_{xd}S_{yd}) /S_{x0}S_{y0}\\
& = \sum_{h,h'}n^h_{ab}(S_{0h})^2S^{h'}_h\sum_{h''}\chi_{h''} \sum_{x,y} S_{xh'}S_{yh''} \delta_{xy} /S_{x0}S_{y0},\\
\end{split}
\end{equation}
where in the third line we have used the Verlinde formula. 
In the last line, we use the fact that the $S$-matrix is symmetric and squares to identity for Virasoro minimal models. Continuing the analysis,
\begin{equation}
\begin{split}
\sum_{c,d}f^{cd}_{ab}~Z_{cd} & = \sum_{h,h'}n^h_{ab}(S_{0h})^2S^{h'}_h\sum_{h''}\chi_{h''} \sum_{x} S_{xh'}S_{h''x} /(S_{x0})^2\\
& = \sum_{h}n^h_{ab}(S_{0h})^2\sum_{h''}\sum_{x}S_{h''x}\chi_{h''}  (\sum_{h'} S_{xh'} S^{h'}_h)/(S_{x0})^2 \\
& = \sum_{h}n^h_{ab}(S_{0h})^2\sum_{h''}\sum_{x}S_{h''x}\chi_{h''}  \delta_{hx}/(S_{x0})^2\\
& = \sum_{h}n^h_{ab}\sum_{h''}S^{h''}_h\chi_{h''}\\
& = \sum_{h,h'}n^h_{ab}S_{h}^{h'}\chi_{h'}.\\
\end{split}
\end{equation}
In the third line we have again used the fact that $S$-matrix is symmetric and squares to identity. The last line is obtained by redefining $h''\rightarrow h'$. We observe that this is exactly \eqref{eq:Z_S}.

\section{Equivalence of disconnected brane configurations for the annulus}
\label{disconnected}

In the bulk of the paper, we considered two geometries dual to a BCFT on the annulus, the high temperature phase corresponding to two disconnected branes in the BTZ black hole background of \eqref{btz}, or the low temperature phase corresponding to a single connected brane in thermal AdS$_3$ with metric \eqref{thermalads}. In \cite{Fujita:2011fp} a third configuration of disconnected branes has been presented in section 3.1 therein. It is the goal of this appendix to show that this putative third solution is just the high temperature BTZ background rewritten in different coordinates.

The putative third configuration is obtained by considering Euclidean AdS$_3$ in Poincar\'e patch coordinates:
\begin{equation}
\label{pmetric}
    ds^2 = \frac{\ell_{AdS}^2}{Z^2}{dT^2 + dX^2 + dZ^2 }.
\end{equation}
A trivial solution for the embedding of an RS brane in this geometry corresponding to a BCFT with planar boundary is given by $X \propto Z$, a half-plane anchored on the $T$-axis, where the constant of proportionality depends on the brane tension. By a conformal transformation, one can obtain from this the solution for the BCFT on a disk $D$ of radius $r_D$. The embedding of the corresponding RS brane is given by the equation
\begin{equation}
\label{pembedding}
    T^2 + X^2 + Z^2 - 2 s_* Z r_D = r_D^2,
\end{equation}
where $s_* = \sinh(\rho_*)$ encodes the brane tension, and $\rho_*$ is as defined in \eqref{eq:universal}. Using the inversion isometry\footnote{Given by $x\rightarrow x/\Xi^2$, $t\rightarrow t/\Xi^2$ and $z\rightarrow z/\Xi^2$, where $\Xi^2\equiv x^2+t^2+z^2$.} of AdS$_3$ one can obtain both solutions where the spacetime removed is the outside or the inside of the disk $D$ (with positive or negative brane tension). It is now easy to put together two such branes $Q$ and $Q'$, of opposite orientations with different disk radii $r_D$ and $r_D'$, in such a way that one keeps only the annulus between the two circular boundaries $\partial D$ and $\partial D'$. In \cite{Fujita:2011fp} this surface was treated as yet another valid configuration for the annulus. Here we want to show that this is just the high temperature BTZ phase in different coordinates.

To see this, first recall that locally the BTZ black hole is just AdS$_3$. Therefore, a coordinate change must exist that takes the Poincar\'e coordinates to the BTZ coordinates in \eqref{btz}. To see this change of coordinates, it is best to connect both coordinate systems to the isometric embedding space $\mathbb{R}^{3,1}$ with coordinates $X_0$, $X_1$, $X_2$, and $X_3$ with metric $dS^2 = - dX_0^2+dX_1^2+dX_2^2+dX_3^2$ in terms of which (Euclidean) AdS$_3$ is given by the hyperboloid of one sheet $X_0^2-X_1^2-X_2^2-X_3^2=\ell_{AdS}^2$. 
For simplicity, we set $\ell_{AdS}=1$ for the remainder of this appendix.
The parametrization of the embedding space coordinates in terms of the Poincar\'e coordinates is well known:
\begin{equation}
    X_0=\frac{1}{2Z} \left(Z^2+X^2+T^2+1\right), \quad X_1 = \frac{T}{Z}, \quad X_2 = \frac{X}{Z}, \quad X_3 = \frac{1}{2Z} \left(Z^2+X^2+T^2-1\right).
\end{equation}
Plugging this into the flat embedding space metric yields \eqref{pmetric}.
The BTZ black hole is given by what in higher dimensions is the hyperbolic slicing of AdS$_3$. The embedding coordinates this time are parametrized as
\begin{eqnarray}
\nonumber
    \frac{X_0}{z_H} &=& \frac{\cosh \frac{x}{z_H}}{z}, \quad \quad \frac{X_1}{z_H} = \sqrt{1-\frac{z^2}{z_H^2}} \frac{\sin \frac{\tau}{z_H}}{z}, \\
    \frac{X_3}{z_H} &=&  \frac{\sinh \frac{x}{z_H}}{z}, \quad \quad \frac{X_2}{z_H} =  \sqrt{1-\frac{z^2}{z_H^2}} \frac{\cos \frac{\tau}{z_H}}{z}.
\end{eqnarray}
This yields the BTZ metric of \eqref{btz} when plugging into the flat embedding space metric.

In the BTZ coordinates, the embedding of a single RS brane associated with the disk on the conformal boundary is given by \cite{Fujita:2011fp}
\begin{equation}
\label{btzembedding}
x(z) = z_H \, \mathrm{arcsinh}( c_0 z)   + x_0
\end{equation}
where $c_0$ encodes the brane tension\footnote{$c_0$ has the same sign as the brane tension.} and $x_0$ is an integration constant. Once again, we can get the annulus by considering the disconnected union of two such branes with different $x_0$, taking care that it is the region between the branes that is kept. Let us start with $x_0=0$. By comparing the expression for $X_3$ in the two parametrizations, we can see that the brane embedding for the disk in BTZ coordinates given by \eqref{btzembedding} in terms of the Poincar\'e coordinates reads
\begin{equation}
   Z^2+X^2+T^2 - 2 Z c_0 = 1
\end{equation}
which is indeed of the form (\ref{pembedding}) with $r_D=1$. To get the general $x_0$ case\footnote{There is only the case with $x_0=0$ in \cite{Fujita:2011fp,Takayanagi:2011zk}, but the $x_0\neq0$ situation is crucial for two separated branes in our BTZ case.}, we need to look at a linear combination of the equation of $X_3$ and $X_0$ to again find the form of (\ref{pembedding}), but this time with $r_D=\cosh(x_0) - \sinh(x_0)$. So, lo and behold, we do indeed find that these putative novel solutions to (\ref{pembedding}) are just the high temperature solutions to (\ref{btzembedding}) in a different coordinate system. The same is obviously true for putting together two disconnected branes of this form separated by a distance $\Delta x=\pi z_0$.

There is another less straightforward way to see that the two disconnected concentric branes are equivalent to the two parallel branes along the meridian in the solid torus representing a BTZ black hole. The ordinary BTZ solid torus is constructed from identifying two ends of a hyperbolic cylinder (possibly with a twist) by an one-generator loxodromic discrete subgroup 
\begin{equation}
    \mathbb{Z}\cong\left\langle
    \begin{pmatrix}
    q & 0\\
    0 & q^{-1}
    \end{pmatrix}
    \right\rangle\subset SL(2,\mathbb{C})
\end{equation}
where $|q|<1$, as in \cite{maloney2010quantum}. The same solid torus can be equivalently constructed from identifying fundamental regions\footnote{The opening angle around $z$-axis is determined by the length of the inner horizon. For a non-rotating BTZ, the regions to be identified are two entire hemispheres.} of two hemispheres centered at the origin in the Poincar\'e upper-space along the radial direction, so that the line segment connecting two North Poles becomes the (outer) horizon, as shown in \cite{Carlip:1994gc,Krasnov:2000zq}. In our case, the two concentric branes $Q$ and $Q'$ centered at $Z=s_*$ in the beginning of this appendix can be viewed as the two hemispheres before identification, and we simply stop there, then it would be brought to the open cylinder cut off by two RS branes in \eqref{btz} via a suitable coordinate transformation, upon which the separation between two North Poles becomes $\Delta x$ right above \eqref{eq:BTZaction}. The center at $s_*$ above (or below) the origin tells if \textit{both} ends of the cylinder are concave (or convex), since $Q$ and $Q'$ have opposite orientations in the Poincar\'e patch \eqref{pmetric}.

\section{Data of both tricritical Ising CFT and BCFT}
\label{app:tri}
In this appendix, out of convenience, we collect results on modular data for the tricritical Ising CFT, and enumerate fusion coefficients for BCFT. We then list all Ishibashi and Cardy states, and their corresponding boundary entropy, as well as partition functions consistent with those boundary conditions.

In the convention of \cite{castro2012gravity}, and in the same order as \eqref{eq:tribasis}, the Virasoro characters of \textit{irreducible} representations $M_{r,s}$ corresponding to six primaries are\footnote{$\chi_{2,3}$, $\chi_{1,3}$ and $\chi_{1,2}$ here are $\chi_{3,2}$, $\chi_{3,1}$ and $\chi_{2,1}$ in \cite{francesco2012conformal}, respectively.}:
\begin{equation}
\label{eq:sixchar}
    \begin{split}
        \chi_{1,1}(\tau)&=K_{1,1}(\tau)-K_{1,-1}(\tau)=\frac{1}{\eta(\tau)}\sum_{n\in\mathbb{Z}}\left(q^{\frac{1}{120}(60n-1)^2}-q^{\frac{1}{120}(60n+11)^2}\right),\\
        \chi_{3,3}(\tau)&=K_{3,3}(\tau)-K_{3,-3}(\tau)=\frac{1}{\eta(\tau)}\sum_{n\in\mathbb{Z}}\left(q^{\frac{3}{40}(20n-1)^2}-q^{\frac{1}{120}(60n+33)^2}\right),\\
        \chi_{2,3}(\tau)&=K_{2,3}(\tau)-K_{2,-3}(\tau)=\frac{1}{\eta(\tau)}\sum_{n\in\mathbb{Z}}\left(q^{\frac{2}{15}(15n-2)^2}-q^{\frac{1}{120}(60n+28)^2}\right),\\
        \chi_{1,3}(\tau)&=K_{1,3}(\tau)-K_{1,-3}(\tau)=\frac{1}{\eta(\tau)}\sum_{n\in\mathbb{Z}}\left(q^{\frac{1}{120}(60n-13)^2}-q^{\frac{1}{120}(60n+23)^2}\right),\\
        \chi_{2,2}(\tau)&=K_{2,2}(\tau)-K_{2,-2}(\tau)=\frac{1}{\eta(\tau)}\sum_{n\in\mathbb{Z}}\left(q^{\frac{1}{30}(30n-1)^2}-q^{\frac{1}{120}(60n+22)^2}\right),\\
        \chi_{1,2}(\tau)&=K_{1,2}(\tau)-K_{1,-2}(\tau)=\frac{1}{\eta(\tau)}\sum_{n\in\mathbb{Z}}\left(q^{\frac{1}{120}(60n-7)^2}-q^{\frac{1}{120}(60n+17)^2}\right),\\
    \end{split}
\end{equation}
where Dedekind's eta function $\eta(\tau)$ is defined as
\begin{equation}
\label{eq:dedekind}
    \eta(\tau)=q^{1/24}\prod_{n=1}^{\infty}(1-q^n),
\end{equation}
where $q=e^{2\pi i\tau}$, and $\tau$ is the modular parameter of the torus.

Their small $q$ expansions are \cite{francesco2012conformal}
\begin{equation}
\label{eq:sixexp}
    \begin{split}
        \chi_{1,1}(q)&=q^{-7/240}(1+q+q^2+q^3+2q^4+2q^5+4q^6+\dots),\\
        \chi_{3,3}(q)&=q^{17/240}(1+q+q^2+2q^3+3q^4+4q^5+6q^6+\dots),\\
        \chi_{2,3}(q)&=q^{137/240}(1+q+2q^2+2q^3+4q^4+5q^5+7q^6+\dots),\\
        \chi_{1,3}(q)&=q^{353/240}(1+q+q^2+2q^3+3q^4+4q^5+6q^6+\dots),\\
        \chi_{2,2}(q)&=q^{1/120}(1+q+2q^2+3q^3+4q^4+6q^5+8q^6+\dots),\\
        \chi_{1,2}(q)&=q^{49/120}(1+q+q^2+2q^3+3q^4+4q^5+6q^6+\dots).\\
    \end{split}
\end{equation}

The modular $S$-matrix in the basis \eqref{eq:tribasis} is 
\begin{equation}
    S=\frac{1}{\sqrt{5}}\begin{pmatrix}
    s_2 & s_1 & s_1 & s_2 & \sqrt{2}s_1 & \sqrt{2}s_2\\
    s_1 & -s_2 & -s_2 & s_1 & \sqrt{2}s_2 & -\sqrt{2}s_1\\
    s_1 & -s_2 & -s_2 & s_1 & -\sqrt{2}s_2 & \sqrt{2}s_1\\
    s_2 & s_1 & s_1 & s_2 & -\sqrt{2}s_1 & -\sqrt{2}s_2\\
    \sqrt{2}s_1 & \sqrt{2}s_2 & -\sqrt{2}s_2 & -\sqrt{2}s_1 & 0 & 0\\
    \sqrt{2}s_2 & -\sqrt{2}s_1 & \sqrt{2}s_1 & -\sqrt{2}s_2 & 0 & 0\\
    \end{pmatrix},
\end{equation}
where $s_1\equiv\sin\left(2\pi/5\right)$ and $s_2\equiv\sin\left(4\pi/5\right)$.

Considering its BCFT, where all $q=e^{2\pi i\tau}$ in \eqref{eq:sixchar}-\eqref{eq:sixexp} are changed into $q=e^{-\pi\delta}$ as in \eqref{eq:Ising_TAdS}, there are six possible boundary conditions in terms of Cardy states written in terms of Ishibashi states:
\begin{equation}
    \begin{split}
        |1\rangle&=C\left[|0\rangle\rangle+\phi|\tfrac{1}{10}\rangle\rangle+\phi|\tfrac{3}{5}\rangle\rangle+|\tfrac{3}{2}\rangle\rangle+\sqrt[\leftroot{-2}\uproot{2}4]{2}|\tfrac{7}{16}\rangle\rangle+\sqrt[\leftroot{-2}\uproot{2}4]{2}\phi|\tfrac{3}{80}\rangle\rangle\right],\\
        |\epsilon''\rangle&=C\left[|0\rangle\rangle+\phi|\tfrac{1}{10}\rangle\rangle+\phi|\tfrac{3}{5}\rangle\rangle+|\tfrac{3}{2}\rangle\rangle-\sqrt[\leftroot{-2}\uproot{2}4]{2}|\tfrac{7}{16}\rangle\rangle-\sqrt[\leftroot{-2}\uproot{2}4]{2}\phi|\tfrac{3}{80}\rangle\rangle\right],\\
        |\sigma'\rangle&=\sqrt{2}C\left[|0\rangle\rangle-\phi|\tfrac{1}{10}\rangle\rangle+\phi|\tfrac{3}{5}\rangle\rangle-|\tfrac{3}{2}\rangle\rangle\right],\\
        |\epsilon\rangle&=C\left[\phi^2|0\rangle\rangle-\phi^{-1}|\tfrac{1}{10}\rangle\rangle-\phi^{-1}|\tfrac{3}{5}\rangle\rangle+\phi^2|\tfrac{3}{2}\rangle\rangle-\sqrt[\leftroot{-2}\uproot{2}4]{2}\phi^2|\tfrac{7}{16}\rangle\rangle+\sqrt[\leftroot{-2}\uproot{2}4]{2}\phi^{-1}|\tfrac{3}{80}\rangle\rangle\right],\\
        |\epsilon'\rangle&=C\left[\phi^2|0\rangle\rangle-\phi^{-1}|\tfrac{1}{10}\rangle\rangle-\phi^{-1}|\tfrac{3}{5}\rangle\rangle+\phi^2|\tfrac{3}{2}\rangle\rangle+\sqrt[\leftroot{-2}\uproot{2}4]{2}\phi^2|\tfrac{7}{16}\rangle\rangle-\sqrt[\leftroot{-2}\uproot{2}4]{2}\phi^{-1}|\tfrac{3}{80}\rangle\rangle\right],\\
        |\sigma\rangle&=\sqrt{2}C\left[\phi^2|0\rangle\rangle+\phi^{-1}|\tfrac{1}{10}\rangle\rangle-\phi^{-1}|\tfrac{3}{5}\rangle\rangle-\phi^2|\tfrac{3}{2}\rangle\rangle\right],
    \end{split}
\end{equation}
where $C\equiv\sqrt{\sin(\pi/5)/\sqrt{5}}$ and $\phi\equiv\sqrt{\cos(\pi/5)}$.

The coefficients in the annulus partition function \eqref{eq:Zab_chi} are: 
\begin{equation}
\begin{split}
n^{0}_{\epsilon\epsilon}= n^{\epsilon'}_{\epsilon\epsilon}=n^{\epsilon}_{\epsilon\epsilon'}=n^{\epsilon''}_{\epsilon\epsilon'}=n^{\epsilon'}_{\epsilon\epsilon''}=n^{0}_{\epsilon'\epsilon'}=n^{\epsilon'}_{\epsilon'\epsilon'}=n^{\epsilon}_{\epsilon'\epsilon''}=n^{0}_{\epsilon''\epsilon''}=1,\\n^{\sigma}_{\epsilon\sigma}=n^{\sigma'}_{\epsilon\sigma}=n^{\sigma}_{\epsilon\sigma'}=n^{\sigma}_{\epsilon'\sigma}=n^{\sigma'}_{\epsilon'\sigma}=n^{\sigma}_{\epsilon'\sigma'}=n^{\sigma}_{\epsilon''\sigma}=n^{\sigma'}_{\epsilon''\sigma'}=1,\\
n^{0}_{\sigma\sigma}=n^{\epsilon}_{\sigma\sigma}=n^{\epsilon'}_{\sigma\sigma}=n^{\epsilon''}_{\sigma\sigma}=n^{\epsilon}_{\sigma\sigma'}=n^{\epsilon'}_{\sigma\sigma'}=n^{0}_{\sigma'\sigma'}=n^{\epsilon''}_{\sigma'\sigma'}=1,\\
n^{h}_{h0}=1,
\end{split}
\end{equation}
and we emphasize again that acoording to \eqref{eq:fusion}, $n^h_{h'h''}$ here is equal to $N^h_{h'h''}$, the fusion coefficients in the Verlinde formula.

Incidentally, the dictionary between primaries and boundary conditions is as follows \cite{Dorey:2009vg} (and \cite{Nepomechie:2001bu,Chim:1995kf} cited therein)\footnote{Note that conventions in \cite{Dorey:2009vg} and \cite{Nepomechie:2001bu} agree, but in \cite{Chim:1995kf} the identifications of $\ket{0+}$ and $\ket{-0}$ to the left column are opposite, and same for $\ket{+}$ and $\ket{-}$.}
\begin{equation}
\begin{split}
    1&:\quad\ket{+},\\
    \epsilon&:\quad\ket{0+},\\
    \epsilon'&:\quad\ket{-0},\\
    \epsilon''&:\quad\ket{-},\\
    \sigma&:\quad\ket{d},\\
    \sigma'&:\quad\ket{0}.\\
\end{split}
\end{equation}
The notations in the right column have their roots in the order parameter $\langle\sigma\rangle$, the expectation value of the leading spin field, on the boundary. ``$+$'', ``$-$'', ``$0$'' respectively mean ``up'', ``down'' and ``zero'' spin, and ``$d$'' stands for ``degenerate'', namely ``$(-0+)$''.

Finally, all possible annulus partition functions are
\begin{equation}
\label{eq:tricritical}
    \begin{split}
        Z_{\epsilon\epsilon}=Z_{\epsilon'\epsilon'}=&\chi_0+\chi_{\epsilon'},\\
        Z_{\epsilon\epsilon'}=&\chi_{\epsilon}+\chi_{\epsilon''},\\
        Z_{\epsilon\epsilon''}=Z_{0\epsilon'}=&\chi_{\epsilon'},\\
        Z_{\epsilon'\epsilon''}=Z_{0\epsilon}=&\chi_{\epsilon},\\
        Z_{\epsilon''\epsilon''}=Z_{00}=&\chi_{0},\\
        Z_{\epsilon\sigma}=Z_{\epsilon'\sigma}=&\chi_{\sigma}+\chi_{\sigma'},\\
        Z_{\epsilon\sigma'}=Z_{\epsilon'\sigma'}=Z_{\epsilon''\sigma}=Z_{\epsilon''\sigma'}=Z_{0\sigma}=&\chi_{\sigma},\\
        Z_{\sigma\sigma}=&\chi_{0}+\chi_{\epsilon}+\chi_{\epsilon'}+\chi_{\epsilon''},\\
        Z_{\sigma\sigma'}=&\chi_{\epsilon}+\chi_{\epsilon'},\\
        Z_{\sigma'\sigma'}=&\chi_{0}+\chi_{\epsilon''},\\
        Z_{0\epsilon''}=&\chi_{\epsilon''},\\
        Z_{0\sigma'}=&\chi_{\sigma'},
    \end{split}
\end{equation}
where we suppressed all of their argument, i.e., the modular parameter $\delta$ of the annulus.

\bibliographystyle{JHEP}
\bibliography{ref}
\end{document}